# Directly visualizing nematic superconductivity driven by the pair density wave in NbSe$_2$

Lu Cao[1], Yucheng Xue[2], Yingbo Wang[2], Fu-Chun Zhang[3], Jian Kang[4], Hong-Jun Gao[2,5], Jinhai Mao[2*], Yuhang Jiang[1*]

[1] College of Materials Science and Opto-Electronic Technology, University of Chinese Academy of Sciences, Beijing 100049, China
[2] School of Physical Sciences, University of Chinese Academy of Sciences, Beijing 100049, China
[3] Kavli Institute for Theoretical Sciences, University of Chinese Academy of Sciences, Beijing 100190, China
[4] School of Physical Science and Technology, ShanghaiTech University, Shanghai 200031, China
[5] Institute of Physics, Chinese Academy of Sciences, Beijing 100190, China

*Correspondence to: jhmao@ucas.ac.cn, yuhangjiang@ucas.ac.cn

**Abstract**

Pair density wave (PDW) is a distinct superconducting state characterized by a periodic modulation of its order parameter in real space. Its intricate interplay with the charge density wave (CDW) state is a continuing topic of interest in condensed matter physics. While PDW states have been discovered in cuprates and other unconventional superconductors, the understanding of diverse PDWs and their interactions with different types of CDWs remains limited. Here, utilizing scanning tunneling microscopy, we unveil the subtle correlations between PDW ground states and two distinct CDW phases — namely, anion-centered-CDW (AC-CDW) and hollow-centered-CDW (HC-CDW) — in 2$H$-NbSe$_2$. In both CDW regions, we observe coexisting PDWs with a commensurate structure that aligns with the underlying CDW phase. The superconducting gap size, $\Delta(\mathbf{r})$, related to the pairing order parameter is in phase with the charge density in both CDW regions. Meanwhile, the coherence peak height, $H(\mathbf{r})$, qualitatively reflecting the electron-pair density, exhibits a phase difference of approximately $2\pi/3$ relative to the CDW. The three-fold rotational symmetry is preserved in the HC-CDW region but is spontaneously broken in the AC-CDW region due to the PDW state, leading to the emergence of nematic superconductivity.

**Introduction**

The condensation of zero center-of-mass momentum Cooper pairs in a superconductor leads to a uniform pairing order parameter in real space. While, a spatially oscillated pairing order parameter emerges when the Cooper pairs possess nonzero momentum. An illustrative example of this effect is the Fulde-Ferrell-Larkin-Ovchinnikov (FFLO) state, which can be realized through the Zeeman effect induced by a magnetic field[1,2]. Recently, a novel state of matter known as the pair density wave[3] (PDW) has been discovered in cuprates[4-6] and other unconventional superconductors[7-10]. The PDW state, characterized by spatial oscillations in pairing order parameter, is driven by electron correlations rather than magnetic field. Intriguingly, the PDW state is often entangled with the charge/spin density wave (C/SDW), offering a fresh insight into the interplay between superconductivity, C/SDWs and other correlated electronic states. Understanding the nature of PDW, including its relation with other ordered states, provides valuable insights into the fundamental physics of correlated electron systems. Despite considerable efforts to uncover the mechanisms driving the emergence of the PDW and its interaction with other quantum states, their relationship remains elusive and demands further research.

In addition to the unconventional superconductors, transition metal dichalcogenide (TMD) materials exhibit abundant ordered states, including CDW, PDW and superconducting phases[11,12]. These TMD materials provide an ideal platform for exploring and unraveling the intricate interplay among these phases. As one of the TMD materials, $2H$-NbSe$_2$ presents a prototypical CDW state below $T_{CDW} \approx 33$ K (Ref.[13]). Furthermore, it exhibits a superconducting (SC) phase below $T_c \approx 7.2$ K, coexisting with the CDW state. Notably, a PDW state, whose electron-pair density has a phase difference of about $2\pi/3$ relative to the CDW phase, was discovered in bulk $2H$-NbSe$_2$ by Josephson STM[12]. Moreover, various symmetry-breaking pairing states have been found in this material, such as two-fold nematic superconductivity in the thin layers due to the potential mixing of different pairing channels or the emergence of topological superconductivity near the upper critical magnetic field[14,15], Ising superconductivity with inversion symmetry-breaking[16,17], and even the unconventional FFLO state with both translational and rotational symmetry-breaking[18]. These phenomena underscore the captivating exotic superconductivity as well as the multiple symmetry-breaking channels in NbSe$_2$.

Spectroscopy imaging scanning tunneling microscopy (SI-STM) has played pivotal role in uncovering the spatial characteristics and local electronic properties. In this study, we utilize the SI-STM to investigate the PDW in $2H$-NbSe$_2$ comprehensively and present an in-depth examination of its interplay with the concurrent CDW. Two types of CDW configurations (AC-CDW and HC-CDW, as defined below) are clearly resolved on the cleaved Se-terminated surface[19,20]. Measured by the tunneling conductance spectra (d$I$/d$V$) and spectroscopy maps, the SC gap size $\Delta(\mathbf{r})$ oscillates in-phase with both CDW orders while the coherence peak height $H(\mathbf{r})$ present a $2\pi/3$ phase difference relative to the CDW orders, demonstrating the existence and universality of the PDW. In HC-CDW region, the PDW state shows a commensurate modulation aligning with the underlying HC-CDW state. In contrast, in the AC-CDW region, the PDW exhibits peculiar modulation with $C_3$ rotational symmetry-breaking, indicating a nematic SC order[21,22] that is not usually observed in other PDW systems[23].

**Results**

**CDW configurations and SC gap modulations in 2$H$-NbSe$_2$**

Each 2$H$-NbSe$_2$ unit cell contains two NbSe$_2$ layers (upper panel of Fig. 1a). After cleavage at the interval between two NbSe$_2$ layers (Fig. 1a), the hexagonal lattice of Se-terminated surface is probed by STM, as shown in the topography $T(\mathbf{r})$ in Fig. 1b. According to the atomic resolution image, a nearly $3a_0$ ($a_0 = 0.344$ nm is the lattice constant of Se-terminated surface) intensity modulation is observed, manifesting the existence of CDW order (Fig. 1b). From the zoomed-in STM topography (Figs. 1c, d), two different types of CDW configurations are identified. The first one has its wave crest (where the largest charge density locates) at one single Se atom in each 3 × 3 unit cell (red dots in Fig. 1c), which is referred to anion-centered-CDW (AC-CDW) following previous study[19]. The other CDW configuration has the charge density peaks at the center of three nearest neighboring Se atoms (yellow dots in Fig. 1d), and is named as hollow-centered-CDW (HC-CDW)[19]. Fourier transforms (FT) show almost the same wave vectors of two CDW configurations with $|\mathbf{Q}_i^{CDW}| \sim 2\pi/3a_0$ (red and yellow dashed circles in Fig. 1e, f). These two CDW configurations form the randomly distributed domains throughout the whole sample and the transitions between different domains are smooth[20] (Fig. 1b and Supplementary Figure 1).

Below $T_c \approx 7.2$ K, the superconductivity emerges and coexists with the CDW states. We first focus on AC-CDW region and measure the d$I$/d$V(V)$ spectra on and off the AC-CDW crest atom as displayed in Fig. 1g. The spectra show the SC coherence peaks around ±1.36 meV accompanied with the hump feature around ±0.60 meV, consistent with previous works on multiple SC gap feature[24-26]. We note the outer SC gap has a well-defined coherence peak feature, which is suited to extract its spatial dependence. However, the inner SC gap has a faint intensity with a broad hump feature at the gap edge or even become dim, hampering the further extraction of accurate SC gap amplitude. The identifiability on the outer coherence peak makes the analyses more objective. Hence, we focus our study on the outer one in this work. In the d$I$/d$V(V)$ spectrum, we define the SC gap size $\Delta$ as half of energy spacing between two coherence peaks and the $H$ as the height of coherence peak (Fig. 1g). Intriguingly, on the AC-CDW crest (yellow curve), the gap size $\Delta$ is larger while the peak height $H$ is lower compared with that off AC-CDW crest (purple curve). These features are more evident in the high-resolution spectra presented in Fig. 1h and 1i. The same phenomena have been repetitively observed in both AC-CDW and HC-CDW regions (Supplementary Figures 2-3), indicating the universal variation of the $\Delta$ and $H$.

**Multiple PDW phases in 2$H$-NbSe$_2$**

To gain insight into the atomic-scale modulation of superconductivity, we conduct the d$I$/d$V$(**r**, $V$) spectra along a specific spatial direction (**Q**$_1$) in the AC-CDW region, as indicated with the orange line in Fig. 2a. In Figs. 2b and 2c, we present a waterfall plot of d$I$/d$V$(**r**, $V$) in the vicinity of two coherence peaks, outlining the modulation of these coherence peaks ($\Delta$) as guided by the black dashed lines. The 1D spatial dependent $\Delta$(**r**) and $H$(**r**) are extracted and presented in Figs. 2d and 2e, respectively (Methods). The corresponding topography $T$(**r**) is also appended for direct comparison (turquoise curve in Figs. 2d and 2e).

The SC gap size $\Delta$(**r**) reaches its maximum at every AC-CDW crest atoms while its minimum is in the AC-CDW trough region. This ~$3a_0$ oscillation could be fitted by the formula $\Delta$(**r**) = $\Delta_0$ + $\Delta_p \cdot \cos(\mathbf{Q}_1 \cdot \mathbf{r})$, where $\Delta_0$ = 1.361 meV, $\Delta_p$ = 0.017 meV and |**Q**$_1$| = $2\pi/3a_0$ (black curve in Fig. 2d). The coherence peak height $H$(**r**) also presents a ~$3a_0$ period modulation (Fig. 2e). However, the maximum of $H$(**r**) has an approximate $2\pi/3$ phase shift relative to the AC-CDW crest (black dashed lines in Fig. 2e). Conventionally, the coherence peak height ($H$) is proportional to the superfluid density[27,28], and has been broadly used to identify the PDW in cuprates. Our measured $2\pi/3$ phase shift between $H$(**r**) and $T$(**r**) indicates a phase difference between the superfluid density and the AC-CDW. Notably, such a $2\pi/3$ phase difference between the superfluid density and CDW is also observed in recent Josephson STM measurements[12], hinting at the same PDW scenario we are focusing on. FTs of $\Delta$(**r**), $H$(**r**) and $T$(**r**) further confirm their spatial oscillations with the same wave vector as the AC-CDW order (Fig. 2f).

When measuring along the other two wave vectors **Q**$_2$ and **Q**$_3$ (rotating by 60° and 120° relative to **Q**$_1$), the d$I$/d$V$(**r**, $V$) spectra indicate the same phenomena in PDW: $\Delta$(**r**) is in-phase with $T$(**r**), but $H$(**r**) has the phase difference of $2\pi/3$ (Supplementary Figures 4-5). The similar phase relationships of $\Delta$ and $H$ have also been observed in HC-CDW region (Supplementary Figure 6), indicating the universality of the PDW state in these two CDW regions. Both the gap modulation and phase difference affirm that we are in the PDW state.

**Rotational symmetry-breaking PDW state in AC-CDW**

To delve further into the correlation between CDW and PDW in 2$H$-NbSe$_2$ system, we collect series of tunneling conductance maps d$I$/d$V$(**r**, $V$) ≡ $g$(**r**, $E = eV$) in atomic scale in the AC-CDW region. Figures 3a-3c depict $g$(**r**, −1.50 meV), $g$(**r**, −1.32 meV) and $g$(**r**, −1.00 meV), which correspond to the difference of local density of states (LDOS) at three energies: |$E$| > $\Delta_{\max}$, |$E$| ≈ $\Delta_0$ and |$E$| < $\Delta_{\min}$, respectively. The corresponding topography $T$(**r**) is presented in Fig. 3d. When |$E$| > $\Delta_{\max}$ (Fig, 3a), the LDOS modulation mainly features the AC-CDW order. For |$E$| ≈ $\Delta_0$ (Fig. 3b), the maximum of LDOS evolves into a triangle lattice similar but is shifted relative to AC-CDW crest atoms. However, at |$E$| < $\Delta_{\min}$ (Fig. 3c), the triangle lattice melts out and more sophisticated atomic modulation appears. This dramatic variation of LDOS observed near the coherence peak, is plausibly related to the intertwined electronic states including CDW, superconductivity and PDW.

Now we focus on the spatial configuration of the PDW in this AC-CDW regions. Centered around each SC gap maximum at AC-CDW crest atoms (white dashed circles in Fig. 3e), three neighboring local SC gap maximum can be identified (indicated by the white arrows), which we assign as the petal structure. These three petals locate along three symmetric directions pointing from Se to its next-nearest neighbor Nb atoms (Supplementary Figure 7). Notably, the intensities of the local SC gaps at these petals have different amplitude distribution which leads to a $C_3$ rotational symmetry-breaking in the AC-CDW region.

To validate the rotational symmetry-breaking induced by the PDW, we track the evolution of the oscillated gap along three high-symmetry lines ($L_1$, $L_2$ and $L_3$ labeled in Fig. 3e) that cutting through these petals. The results are presented in Fig. 3f. Along these three directions, the spatial variations of the SC gap all exhibit periodic modulation behavior. Moreover, it is evident that the gap size along $L_3$ (yellow curve) has a clearly larger amplitude than those along the other two directions ($L_1$ and $L_2$). This suggests a three-fold rotational symmetry-breaking induced by the PDW due to the intra-structure, *i.e.*, the petals. We further slightly translate the spatial positions of $L_1$, $L_2$ and $L_3$ to prove the robustness of their choice. Details are shown in Supplementary Note 2 and Supplementary Figure 14. To confirm that the observed features in Fig. 3e, we statistically average the local SC gap size on all petals. Specifically, we count the gap size along three symmetric directions ($D_1$, $D_2$ and $D_3$) marked by colored arrows radiating from each AC-CDW crest atom (one example shown in the inset of Fig. 3g). To quantitatively assess the anisotropy of gap distribution, we average these gap values for each direction to reduce the randomness. Consistently, the averaged gap size along the yellow arrow direction ($D_3$) is significantly larger at the distance of 0.35 nm away from crest atom.

To provide further affirmation for the conclusion above, in Fig. 3e, more accurate areas should be calculated. We select the petals around each crest atom into three-fold symmetric areas: $A_1$, $A_2$ and $A_3$ (examples shown in the insets of Fig. 3h), and calculate the local averaged gap size for each area (Methods). By statistical analysis, we generate histograms of these averaged gap sizes for all the $A_1$, $A_2$ and $A_3$ areas around every crest atom in Fig.3e, and these histograms are plotted in Fig. 3h. The three dashed vertical lines there indicate the overall averaged values for the $A_1$, $A_2$ and $A_3$ respectively. And the black arrows indicate the ±standard deviation, which are smaller than the variations of SC gap size. In agreement with above two methods, areas $A_3$ has the largest averaged gap size, with approximately 1.5% larger than that in area $A_1$ and $A_2$. To further strengthen the conclusion above, we shift the chosen areas $A_1$, $A_2$ and $A_3$ in ±x and ±y directions for 1 pixel (Supplementary Note 3). The results in Supplementary Tables 1-3 show the insensitivity of averaged value or standard deviation before and after shift. We note there might be anisotropic tip effect or artifact signal from topography. Thus, we carry out rigorous processes to rule out the tip anisotropy and artifact signal from topography. Details are described in Supplementary Notes 1, 5-6, and Supplementary Figures 12-13, 16-17. In conclusion, the $C_3$ rotational symmetry breaking of SC gap size in AC-CDW region is solid.

In addition, we use the same method to extract the LDOS intensity in AC-CDW region in Fig. 3a-3c and present the histograms in Supplementary Figure 8. When $|E| > \Delta_{max}$ (Supplementary Figure 8b), the histogram distribution and the averaged LDOS (indicated by dashed lines) are nearly identical in $A_1$, $A_2$ and $A_3$. However, when $|E| \leq \Delta_0$ (Supplementary Figure 8d, f), the histogram distribution and the averaged LDOS in $A_3$ is significantly lower than other two, providing further evidence of three-fold symmetry-breaking. The 2D coherence peak height map $H(\mathbf{r})$ is also extracted and presented in Supplementary Figure 8g, which is nearly identical to $g(\mathbf{r}, -1.32 \text{ meV})$ (Fig. 3b). And the corresponding intensity histogram of $H(\mathbf{r})$ is shown in Supplementary Figure 8h. Similar to Supplementary Figure 8d, the averaged peak height in $A_3$ is smaller than that in $A_1$ and $A_2$. Our results suggest that the rotational symmetry-breaking is strongly linked to superconductivity, as it emerges solely within the SC gap regime. These gap and LDOS anisotropic distributions analyzed above suggest that the PDW in AC-CDW region leads to a spontaneously rotational symmetry-breaking. It is worth noting that the rotational symmetry-breaking in this regime imparts electronic nematicity to both the PDW and SC phase in bulk $NbSe_2$.

Importantly, similar rotational symmetry-breaking states induced by the PDW along the other two directions ($\mathbf{L}_1$, $\mathbf{L}_2$) are also observed at other locations (Supplementary Figures 9-10). Our data may be explained by the scenario that the PDW in $NbSe_2$ are constructed by three components with equal weight or intensity but rotated 120° relatively. The interplay between charge order and PDW strengthen one unique component that leads to the three-fold symmetry-breaking. This is reminiscent of the nematic order[29] or superconductivity[30] in twisted bilayer graphene and electronic nematicity in the Kagome superconductor[31,32], where the interaction plays a dramatic role on the three-fold symmetry-breaking.

To further corroborate the existence of nematic SC phase, we provide a result in Fig. 4, where two symmetry breaking regions (r1 and r2) coexist in one gap map (Fig. 4b). The green and white dashed lines highlight the domain boundaries. The symmetry breaking components are highlighted by green and white arrows, respectively. It can be found that the $C_3$ symmetry breaking in these two regions has different directions. In addition, we use the same statistics method as described above to quantitatively confirm the $C_3$ rotational symmetry breaking within the SC gap map (Fig. 4c-d). Notably, the directions of $C_3$ symmetry breaking between the two domains show obvious differences. These results convincingly demonstrate that the observed symmetry breaking effect is genuine rather than anisotropy or artifact caused by tip.

### $C_3$ symmetric PDW state in HC-CDW region

We now turn to discussion of the PDW states in the HC-CDW region. Here, we discover that even a minor change in the CDW format can significantly alters the PDW configurations. The map of the SC gap size, $\Delta(\mathbf{r})$, extracted for HC-CDW (topography in Fig. 5a) is displayed in Fig. 5b, revealing that the PDW state maintains the same periodicity of ~$3a_0$. In contrast to the PDW configuration at AC-CDW regime, in this case, the maximum gap size precisely aligns with the three brightest Se atoms without any additional petal distribution. We subsequently apply the previously described methods to detect any potential anisotropic gap distribution and present the results in Fig. 5c-e. Remarkably, the gap sizes along all three symmetric lines ($\mathbf{L}_1$, $\mathbf{L}_2$ and $\mathbf{L}_3$) have nearly the identical amplitudes in Fig. 5c, devoid of any evident three-fold symmetry-breaking features. This result is further supported by the averaged $\Delta(\mathbf{r})$ along all three arrowed directions ($\mathbf{D}_1$, $\mathbf{D}_2$ and $\mathbf{D}_3$) which exhibit a consistent distribution (Fig. 5d). Furthermore, the histograms as well as the overall averaged values of the local gap size in three areas $A_1$, $A_2$ and $A_3$ show no indication of $C_3$ rotational symmetry-breaking (Fig. 5e). These phenomena are repeatable in other HC-CDW regions (Supplementary Figure 11). Therefore, we can conclude that the PDW state in HC-CDW region still preserves the three-fold symmetry, which stands in stark contrast to the PDW behavior in AC-CDW region.

The existence of distinct PDWs in these two CDW regions underscores the subtle interaction or dependency between CDW and PDW. Previous theoretical calculations, which considered the electronic interactions, lattice dynamics, and the energetics of both phases, aim to comprehend the underlying physics of the interplay between PDW and CDW[33,34]. In this context, NbSe$_2$ offers a unique opportunity to study the intertwined CDW and PDW with different configurations. The STM topography reveals that the two CDWs have the same periodicity and symmetry but exhibit different phases relative to the underlying NbSe$_2$ lattice. This weak variation in CDW configuration significantly impacts the interplay between PDW and CDW. One plausible explanation is that the CDW configuration slightly alters the local NbSe$_2$ lattice due to imperfect commensurability, which eventually influences the lattice dynamics that govern their interaction and the ground state of the PDW.

Intriguingly, although the PDW states in AC- and HC-CDW regions are two distinct phases, the $\Delta(\mathbf{r})$ evolves continuously at their boundary (Fig. 5f, g), suggesting that the transition between two PDW phases can occur at the atomic scale with smoothness (Fig. 5h). These findings highlight the complex interplay among multiple CDW, superconductivity and PDW phases in this system.

**Discussion**
The CDW and PDW patterns are highly overlapped in HC-CDW regions, suggesting a positively correlation between them, *i.e.*, both the maximum of C/PDW locates at the three nearest Se atoms and preserve the $C_3$ rotational symmetry. In contrast, in AC-CDW regions, the emergence of the petal structure in PDW pattern is unexpected and cast the strong evidence about the complex interplay between the AC-CDW and superconductivity, since only one Se atom is brightest from AC-CDW while the PDW has $C_3$ petal pattern around this Se atom. The $C_3$ symmetry breaking of the PDW further indicates the relation between CDW, superconductivity and PDW is entangled in NbSe$_2$.

Our STM measurements demonstrate multiple PDW phases in 2*H*-NbSe$_2$, *i.e.*, $C_3$ symmetric one in HC-CDW region and nematic one in AC-CDW region. It is imperative to investigate the driving force of the $C_3$ rotational symmetry-breaking in the nematic PDW. In many instances, rotational symmetry-breaking can be induced by local random strain[35,36]. However, $C_3$ symmetry-breaking is absent from both lattice and CDW at the same locations (Supplementary Figure 12). Moreover, PDWs in all studied HC-CDW regions still preserve $C_3$ symmetric, further ruling out the strain as the driving force of the nematicity. As we discussed earlier, in other AC-CDW regions, we also observe larger gap amplitudes along **D**$_1$ or **D**$_2$ direction compared to the other two directions (Supplementary Figures 9-10), providing the additional support for the hypothesis that spontaneous rotational symmetry-breaking is driven by an intrinsic electronic instability within the pairing channel[14,15,37]. While the PDW has been reported in NbSe$_2$ by Josephson STM[12], our work reveals the unexpected occurrence of nematic superconductivity in the AC-CDW region, which signifies a specific configuration of the SC anisotropy. Comparing to recently reported PDW works in unconventional superconductors[4-10], 2*H*-NbSe$_2$ exhibits rich CDW and PDW phases that provide us an ideal platform to study their interplay. Of particular significance, our finding indicates that different PDW phases could be realized within one material through the CDW form selection, which offers the possibility on engineering the PDW by manipulating the CDW phases[38-40].

## Methods
### STM/S experiments
High quality 2$H$-NbSe$_2$ crystals are purchased from HQ Graphene company. The crystals are cleaved in ultra-high vacuum and transferred *in-situ* into the STM head immediately. All experiments are conducted at 300 mK (Unisoku). Pt-Ir tips are used in the measurements. The STM topography images are acquired under the constant-current mode. And the standard lock-in amplifier is utilized to obtain the d$I$/d$V$ spectra, spectra line-cuts and maps with $V_{mod}$ = 0.05 mV and $f$ = 773 Hz. The energy resolution calibration at 300 mK is described in Supplemental Note 4 and Supplemental Figure 15.

### Extraction of SC gap size $\Delta$ and coherence peak height $H$
We first define the SC gap size $\Delta$ as half of energy spacing between two coherence peak maximums and the $H$ as the d$I$/d$V$ intensity of coherence peak maximum[28]. In order to obtain a more accurate SC gap value, we fit the curve by five-order polynomial $g(V)$ = $a_1V^5+a_2V^4+a_3V^3+a_4V^2+a_5V+a_6$, with confidence level 95%. Then, we utilize the ordinary peak detection algorithm to find the peak maximum and extract the SC gap size $\Delta$ as well as coherence peak height $H$.

### Calculating the local averaged SC size in PDW patterns
In order to calculating the local averaged SC size $\Delta$ in Fig. 3e, we choose an area which contains 16 pixels to form a grid ($A_3$ in the inset of Fig. 3h). The area $A_3$ is designed to match the shape of the local SC gap distribution. By rotating 120° and 240°, we get area $A_1$ and $A_2$, respectively (inset of Fig. 3h). In $A_3$, the averaged value within the 16 pixels is defined as the local SC gap averaged value. In $A_1$ and $A_2$, due to the mismatch between the grid-pixel and the data-pixel, we divide each grid-pixel into 10 × 10 sub-grid-pixel. Each sub-grid-pixel has an accurate value because it has a complete overlap with the underlying data-pixel. Thus, the averaged SC gap size can be calculated in $A_1$ and $A_2$. Next, by statistics, we obtain the histograms of the local averaged SC size $\Delta$ in area $A_1$, $A_2$ and $A_3$ as shown in Fig. 3h. Finally, we calculate the overall averaged value for $A_1$, $A_2$ and $A_3$, as marked by the vertical dashed lines in Fig. 3h. The ±standard deviation are indicated by black arrows. The histograms in Fig. 5e are obtained by same method but with slightly different grid to match the local SC gap distribution. We also use this method to count up the LDOS (Supplementary Figure 8) and the gap size at other locations (Supplementary Figures 9-11).

### Data availability
The data that support the findings of this work are available within the main text and Supplementary Information. The data are available from the corresponding authors upon request.

### Code availability
The code used to analyze the data in this work is available from the corresponding authors upon request.

**Acknowledgements**
We thank Y. Zhang and K. Jiang for helpful discussion. We also thank J. Yan and H. Zhao for technical assistance. The work is supported by National Key R&D Program of China with Grants 2019YFA0307800 (J.M.), National Natural Science Foundation of China with Grants 11974347 (J.M.), 12074377 (Y.J.), the China Postdoctoral Science Foundation with Grant No. 2022M723111 (L.C.), the Fellowship of China National Postdoctoral Program for Innovative Talents with grant No. BX20230358 (L.C.) and the Fundamental Research Funds for the Central Universities (L.C., J.M. and Y.J.).

**Author contributions**
Y.J., J.M. and L.C. conceived and designed the experiments. L.C. and Y.X. performed the STM experiments. L.C., Y.W., H.-J.G., J.M. and Y.J. analyzed the raw data and plotted the figures. L.C., F.-C.Z., J.K., J.M. and Y.J. wrote the manuscript with inputs from all authors. Y.J. and J.M. supervised the project.

**Competing interests**
The authors declare no competing interests.

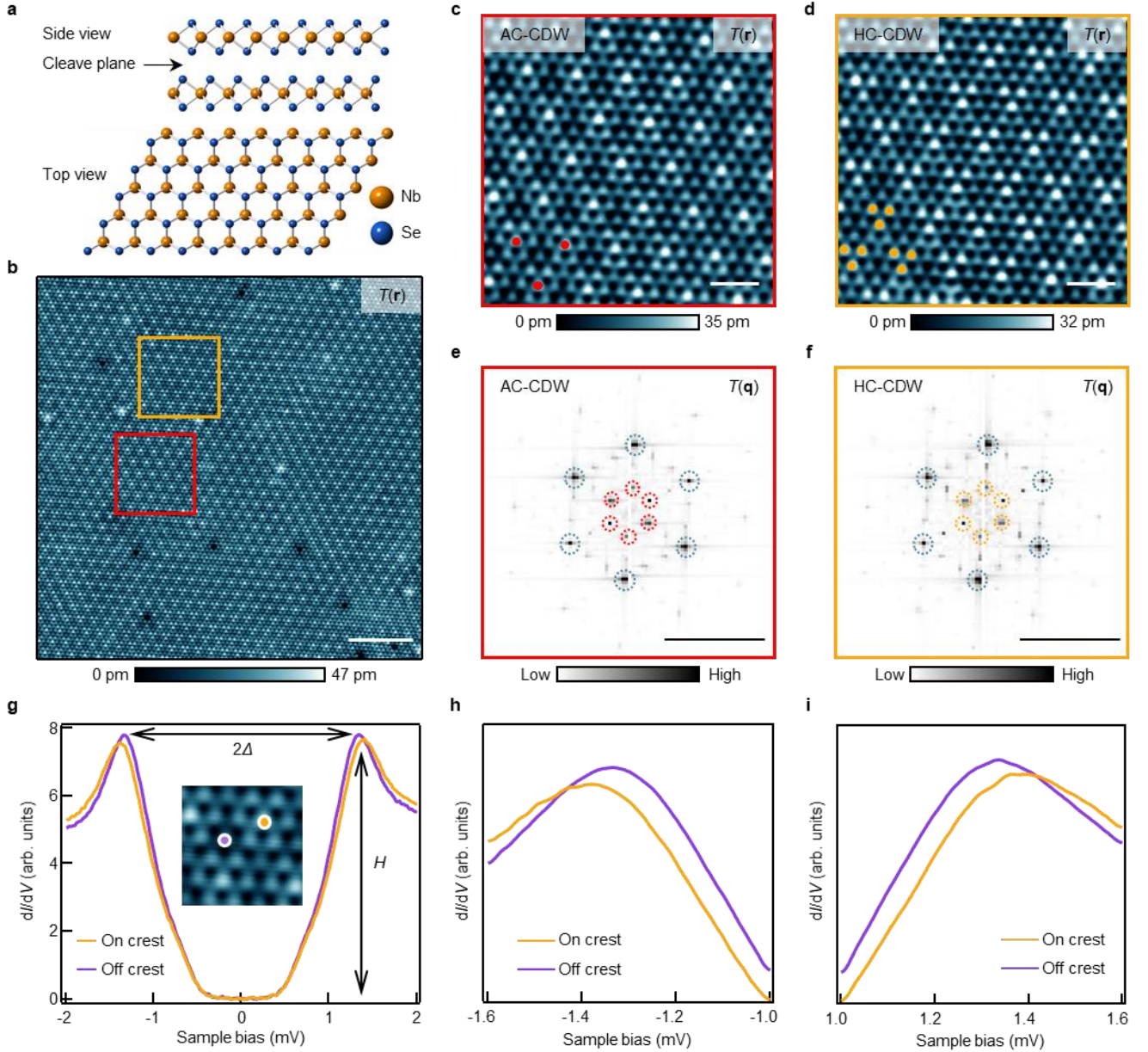

**Fig. 1| CDW configurations and SC gap modulations in 2$H$-NbSe$_2$. a** The side view (upper panel) and top view (lower panel) of lattice structure of 2$H$-NbSe$_2$. **b** The atomic resolution in topography $T(\mathbf{r})$, showing the Se-terminated surface after cleavage. The typical CDW order with period ~3$a_0$ is clearly observed, where $a_0$ = 0.344 nm is the Se lattice constant. Scale bar: 5 nm. **c** The zoomed-in STM topography $T(\mathbf{r})$ of AC-CDW as marked by the red square in (**b**). Only one brightest Se atom (red dots) appears in each AC-CDW unit cell. The AC-CDW crest is located at the brightest Se atom. Scale bar: 1 nm. **d** The zoomed-in STM topography $T(\mathbf{r})$ of HC-CDW as marked by the yellow square in (**b**). Three Se atoms (yellow dots) with the same brightness appear in each HC-CDW unit cell. The HC-CDW crest is located at the center of these three Se atoms. Scale bar: 1 nm. **e,f** Fourier transform images $T(\mathbf{q})$ of (**c**) and (**d**), showing the lattice Bragg peaks (turquoise dashed circles) and the CDW wave vectors $|\mathbf{Q}_i^{CDW}|$ ~2$\pi$/3$a_0$ (i = 1, 2, 3. Red/yellow dashed circles). Scale bar: 0.5 Å$^{-1}$. **g** The d$I$/d$V(V)$

spectra taken on (yellow curve) and off (purple curve) the AC-CDW crest as marked by the dots in the inset. Here, the SC gap size $\Delta$ is defined by the half energy spacing between two coherence peaks and the $H$ is defined as the coherence peak height. **h,i** The zoomed-in spectra around the coherence peaks, showing the larger $\Delta$ and lower $H$ on AC-CDW crest (yellow curves) compared to that off AC-CDW crest (purple curves). Experimental setpoints in (**b-d**) and (**g-i**): $V_s = -4$ mV, $I_t = 400$ pA.

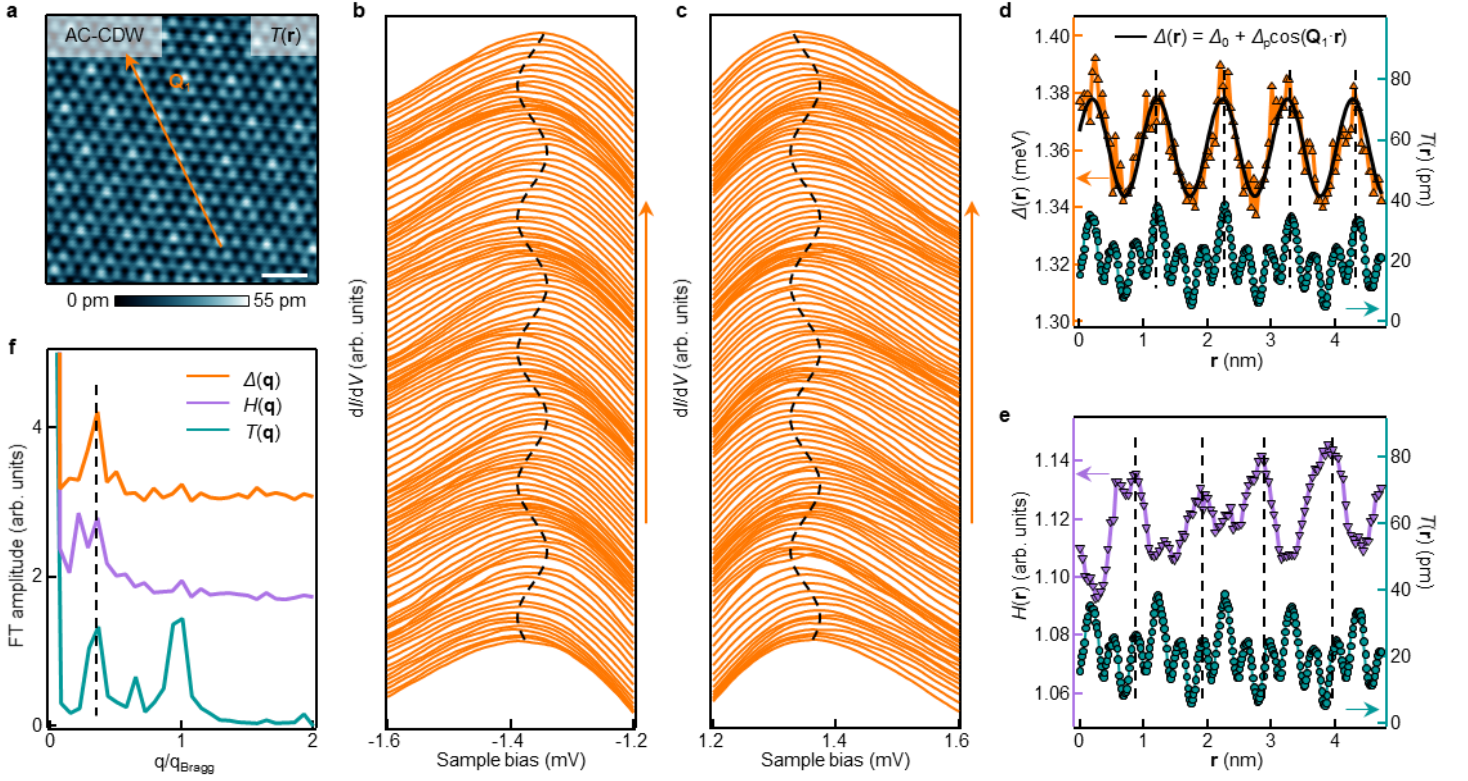

**Fig. 2| PDW state in 2$H$-NbSe$_2$. a** A small-scale topography $T(\mathbf{r})$ of AC-CDW order. Scale bar: 1 nm. **b,c** The waterfall plot of d$I$/d$V(V)$ tunneling spectra taken along the orange arrow marked in (**a**). Black dashed lines are guides to the eye to follow the modulation of coherence peaks. **d** 1D gap map $\Delta(\mathbf{r})$ (orange curve) extracted from (**b**) and (**c**). The $T(\mathbf{r})$ (turquoise curve) presents the modulation of topography at AC-CDW region. The black curve is fitted by using the formula $\Delta(\mathbf{r}) = \Delta_0 + \Delta_\mathrm{p}\cos(\mathbf{Q}_1\cdot\mathbf{r})$, where $\Delta_0 = 1.361$ meV, $\Delta_\mathrm{p} = 0.017$ meV, $|\mathbf{Q}_1| = 2\pi/3a_0$. The $\Delta(\mathbf{r})$ is in-phase with CDW order (black dashed lines). **e** $H(\mathbf{r})$ (purple curve) extracted from (**b**) and (**c**). The $T(\mathbf{r})$ (turquoise curve) presents the modulation of topography at AC-CDW region. The phase difference between the $H(\mathbf{r})$ and CDW order is about $2\pi/3$, as indicated by the black dashed lines. **f** Fourier transforms (FT) of $\Delta(\mathbf{r})$, $H(\mathbf{r})$ and $T(\mathbf{r})$ in (**d**) and (**e**), labelled as $\Delta(\mathbf{q})$, $H(\mathbf{q})$ and $T(\mathbf{q})$, respectively. It shows that $\Delta(\mathbf{r})$ and $H(\mathbf{r})$ have the same period of $T(\mathbf{r})$ (black dashed line). Offsets are set to orange and purple curves for clarity. Experimental setpoints in (**a-c**): $V_\mathrm{s} = -4$ mV, $I_\mathrm{t} = 400$ pA.

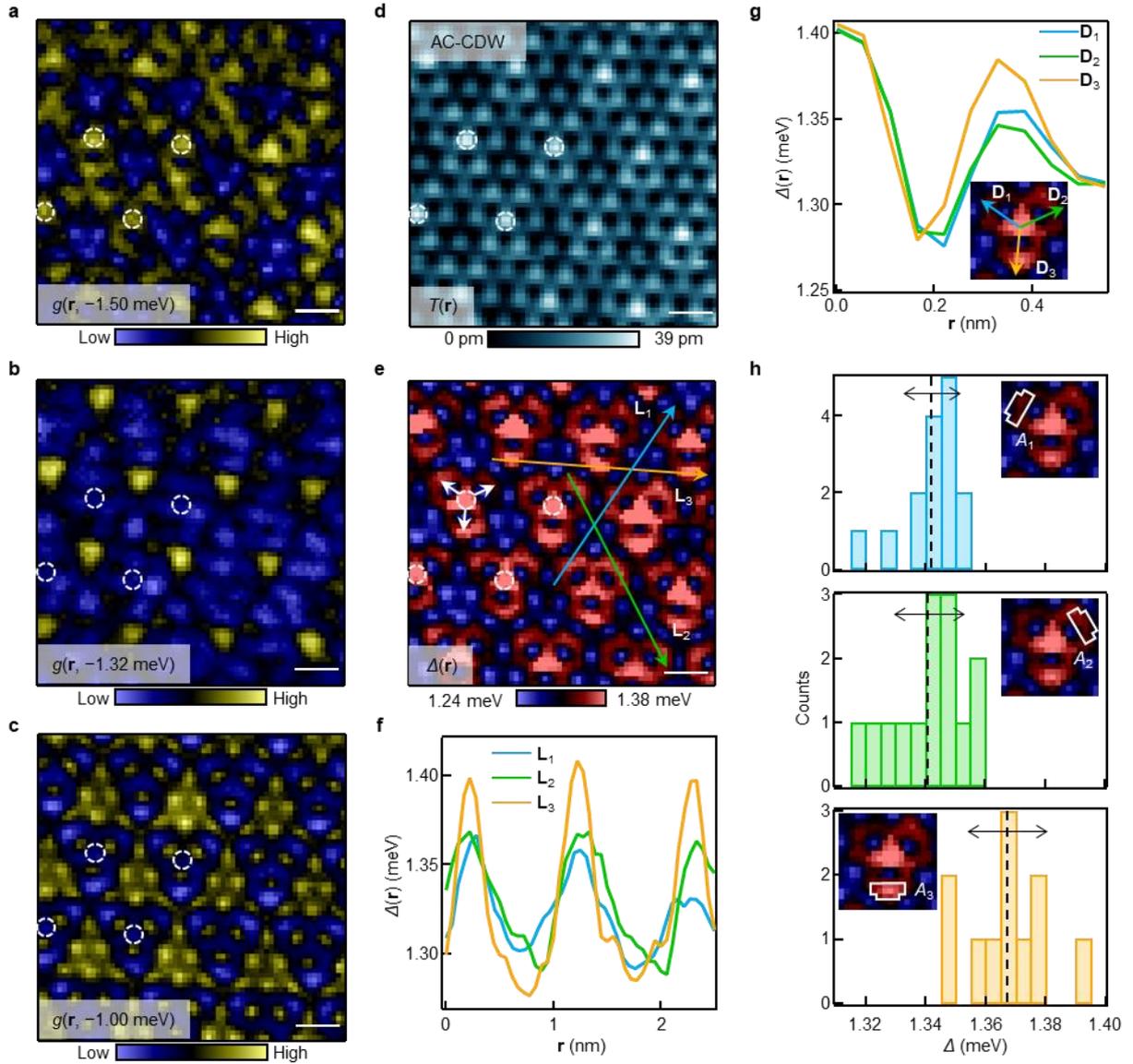

**Fig. 3| PDW state with $C_3$ rotational symmetry-breaking in AC-CDW region. a-c** Simultaneously obtained tunneling spectra maps $dI/dV(\mathbf{r}, V) \equiv g(\mathbf{r}, E = eV)$ taken at −1.50, −1.32, −1.00 meV respectively, showing the quite different LDOS in such small energy interval. Scale bar: 0.5 nm. **d** The corresponding topography $T(\mathbf{r})$ of AC-CDW order in (**a-c**). Scale bar: 0.5 nm. **e** SC gap map $\Delta(\mathbf{r})$ extracted from tunneling spectra maps. Scale bar: 0.5 nm. White dashed circles in (**a-e**) mark the AC-CDW crest configuration. **f** The $\Delta(\mathbf{r})$ line-profiles taken along three symmetric lines, $L_1$, $L_2$ and $L_3$ as marked in (**e**). The $\Delta(\mathbf{r})$ amplitude along $L_3$ is clearly larger compared to other two directions ($L_1$ and $L_2$). **g** The averaged $\Delta(\mathbf{r})$ along three symmetric directions ($D_1$, $D_2$ and $D_3$ in the inset) extracted from all the crests in (**e**). **h** Histograms of local averaged SC gap size $\Delta$ in areas $A_1$, $A_2$ and $A_3$ around all crests in (**e**). An example of the definition of $A_1$, $A_2$ and $A_3$ is shown in the inset. Vertical dashed lines indicate the overall averaged gap value for $A_1$, $A_2$ and $A_3$, respectively. Black arrows indicate the ±standard deviation. The results in (**f-h**) all show the $C_3$ rotational symmetry-breaking of PDW state. Experimental setpoints in (**a-d**): $V_s$ = −4 mV, $I_t$ = 400 pA.

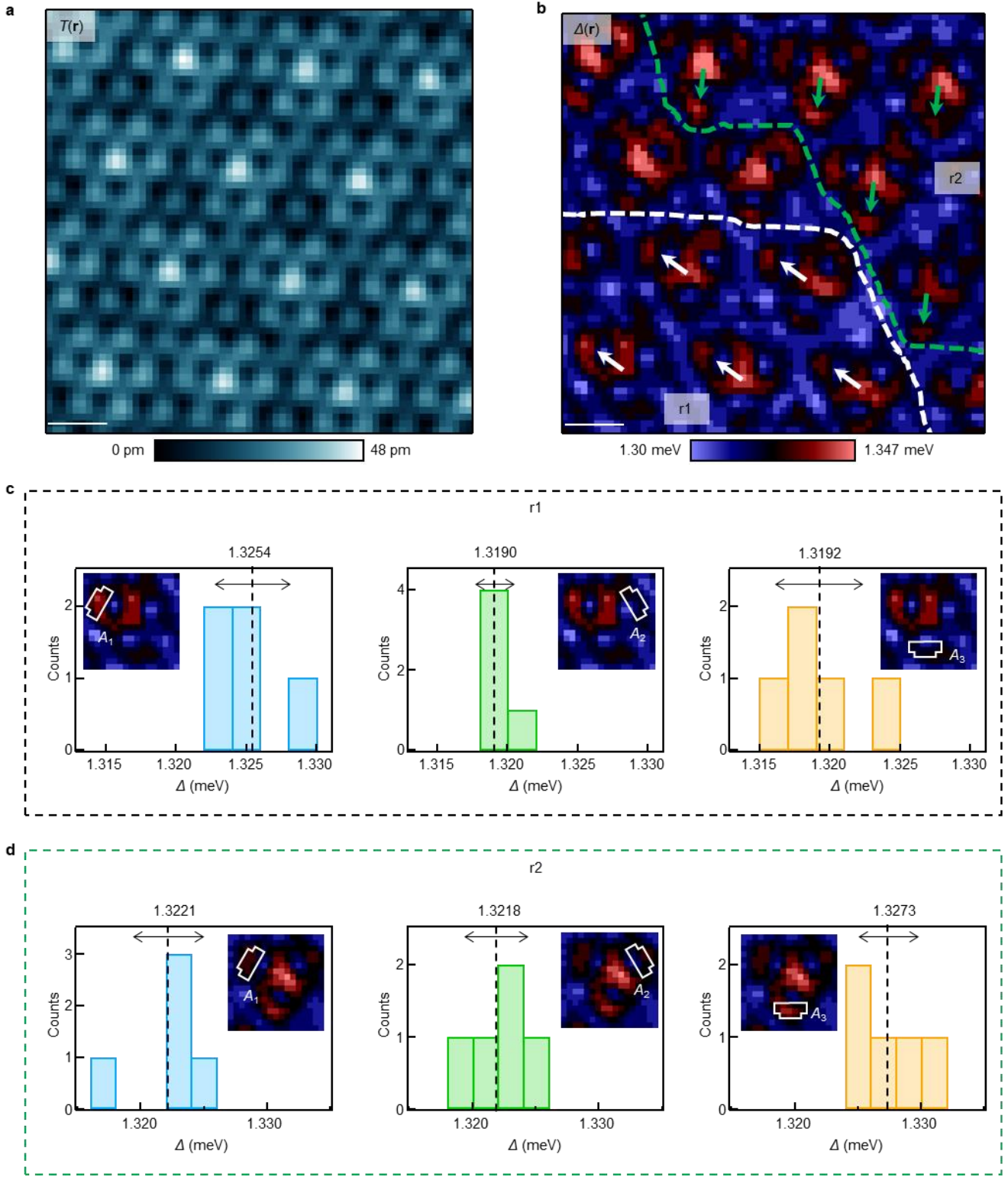

**Fig. 4| Gap map with $C_3$ symmetry breaking along different directions. a**, The topography $T(\mathbf{r})$ of AC-CDW region. Scale bar: 0.5 nm. **b**, The corresponding SC gap map $\Delta(\mathbf{r})$. There are two regions (r1 and r2) separated by dashed lines. The petals exhibiting the largest SC gaps, which lead the $C_3$ rotational symmetry breaking, are denoted by arrows emanating from their centers. Scale bar: 0.5 nm. **c**, Histograms of local averaged SC gap size $\Delta$ in areas $A_1$, $A_2$ and $A_3$ within region r1. **d**, Histograms of local averaged SC gap size $\Delta$ in areas $A_1$, $A_2$ and $A_3$ within region r2. In (**c**) and (**d**), an

example of the definition of $A_1$, $A_2$ and $A_3$ is shown in the inset. Vertical dashed lines and numbers indicate the overall averaged gap value for $A_1$, $A_2$ and $A_3$, respectively. Black arrows indicate the ±standard deviation. Experimental setpoints in (**a-b**): $V_s = -4$ mV, $I_t = 400$ pA.

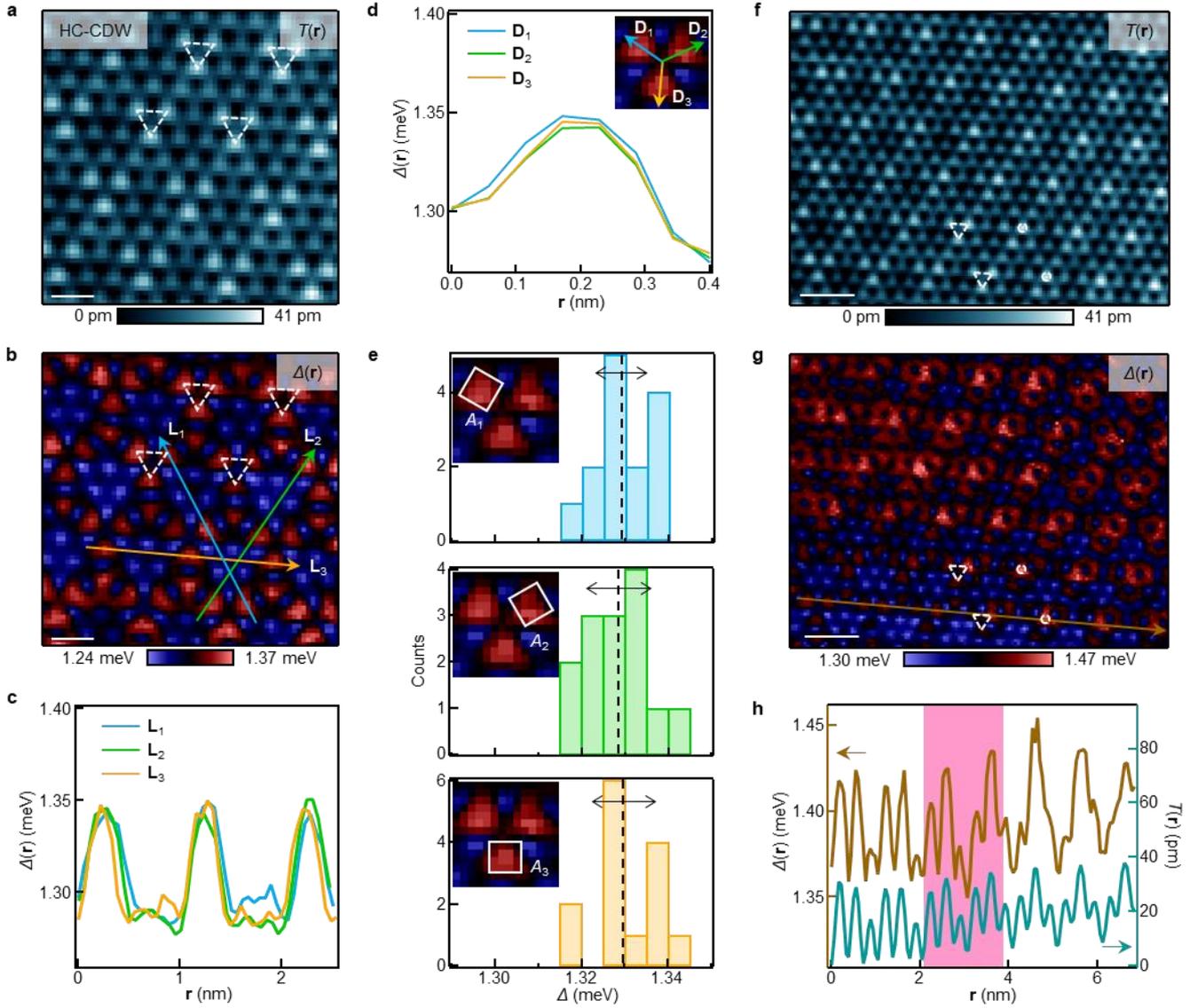

**Fig. 5| PDW state in HC-CDW region and the transition boundary. a** Topography $T(\mathbf{r})$ of HC-CDW order. White dashed triangles mark the HC-CDW configuration. The crest in HC-CDW region is defined as the center of the white dashed triangles. Scale bar: 0.5 nm. **b** SC gap map $\Delta(\mathbf{r})$ in HC-CDW region. Scale bar: 0.5 nm. **c** The $\Delta(\mathbf{r})$ line-profiles taken along three symmetric lines, $L_1$, $L_2$ and $L_3$, around the crest as marked in (**b**). The $\Delta(\mathbf{r})$ amplitude along three lines are almost same. **d** The averaged $\Delta(\mathbf{r})$ along three symmetric directions ($D_1$, $D_2$ and $D_3$ in the inset) extracted from all the crest in (**b**). **e** Histograms of local averaged SC gap size $\Delta$ in areas $A_1$, $A_2$ and $A_3$ around all crest centers in (**b**). Vertical dashed lines indicate the overall averaged gap value for $A_1$, $A_2$ and $A_3$, respectively. Black arrows indicate the ±standard deviation. The results in (**c-e**) show PDW state preserves the $C_3$ rotational symmetry in HC-CDW region. **f** Topography $T(\mathbf{r})$ contains both AC-CDW and HC-CDW orders and a continuous transition boundary. Scale bar: 1 nm. **g** SC gap map $\Delta(\mathbf{r})$ from spectra maps in the transitional region. Scale bar: 1 nm. **h** The 1D extracted $\Delta(\mathbf{r})$ along the brown line in (**g**). The corresponding $T(\mathbf{r})$ is also appended (turquoise curve). The pink bar is the

position of transition boundary. White dashed circles and triangles in (**f**, **g**) mark the AC- and HC-CDW configuration. Experimental setpoints: $V_s = -4$ mV, $I_t = 400$ pA.